\documentclass[10 pt]{article}

\usepackage{amssymb}
\usepackage{lineno}

\usepackage{comment}
\usepackage{amsmath } 
\usepackage{amssymb }
\usepackage{amsthm }  
\usepackage{physics}
\usepackage{graphicx}
\usepackage{wrapfig} 
\usepackage{dsfont}
\usepackage{bm}
\usepackage{hyperref}
\usepackage{xcolor}
\usepackage{soul}
\usepackage[margin = 1. in]{geometry}

\newcommand\pd{\partial}
\newcommand\diff{\text{d}}

\newcommand\fL{\mathcal{L}}

\newcommand\Indic{\mathds{1}}

\newcommand\mat[1]{\begin{pmatrix} #1 \end{pmatrix}}
\newcommand\avg[1]{\left\langle #1 \right\rangle}
\newcommand\avgph[1]{\avg{#1}_{\hat{p}}}
\newcommand\stiv{STIV }
\newcommand\noneq{\text{neq}}
\newcommand\tot{\text{tot}}
\newcommand\N{\mathcal{N}}

\begin{document}

\begin{center}
    \LARGE
    A statistical mechanics derivation and implementation of non-conservative phase field models for front propagation in elastic media \\ 
    \vspace{1 cm}
    \large
    Travis Leadbetter$^\text{a}$, Prashant K. Purohit$^\text{b}$, and Celia Reina$^\text{b,1}$\\
    September 2023\\
    \vspace{1 cm}
    \footnotesize
    {}$^\text{a}$Graduate Group in Applied Mathematics and Computational Science, University of Pennsylvania, Philadelphia, PA, 19104;\\
    {}$^\text{b}$Department of Mechanical Engineering and Applied Mechanics, University of Pennsylvania, Philadelphia, PA, 19104;\\{}$^\text{1}$To whom correspondence should be addressed: creina@seas.upenn.edu
\end{center}

\begin{abstract}
    Over the past several decades, phase field modeling has been established as a standard simulation technique for mesoscopic science, allowing for seamless boundary tracking of moving interfaces and relatively easy coupling to other physical phenomena.
However, despite its widespread success, phase field modeling remains largely driven by phenomenological justifications except in a handful of instances. 
In this work, we leverage a recently developed statistical mechanics framework for non-equilibrium phenomena, called Stochastic Thermodynamics with Internal Variables (STIV), to provide the first derivation of a phase field model for front propagation in a one dimensional elastic medium without appeal to phenomenology or fitting to experiments or simulation data. 
In the resulting model, the variables obey a gradient flow with respect to a non-equilibrium free energy, although notably, the dynamics of the strain and phase variables are coupled, and while the free energy functional is non-local in the phase field variable, it deviates from the traditional Landau-Ginzburg form. 
Moreover, in the systems analyzed here, the model accurately captures stress induced nucleation of transition fronts without the need to incorporate additional physics.
We find that the STIV phase field model compares favorably to Langevin simulations of the microscopic system and we provide two numerical implementations enabling one to simulate arbitrary interatomic potentials. 
\end{abstract}

\section*{Highlights}
\begin{itemize}
    \item Non-conservative phase field models are derived without phenomenology or parameter fitting. 
    \item Accurate prediction of mechanical and thermodynamic quantities.
    \item Arbitrary interaction potentials and external driving.
\end{itemize}

\section*{Keywords}
Allen-Cahn | Gradient Flow | Coarse Graining | Internal Variables | STIV | Gradient Flow | GENERIC

\section{Introduction}
\label{sec:Intro}
Predictive mesoscale models are important for applications in engineering, materials science, biology, and medicine. 
One of the most successful methods, particularly for problems with moving boundaries between regions of differing microscopic character, is that of phase field modeling \cite{chen2002,steinbach2009phase,moelans2008introduction,emmerich2003diffuse}. 
Phase field models have been highly successful in describing solid-solid and solid-liquid phase transformations \cite{steinbach2006multi,boettinger2002phase,echebarria2004quantitative}, crack propagation \cite{miehe2010phase,spatschek2011phase,bourdin2014morphogenesis}, crystal growth \cite{kobayashi1993modeling,meca2013phase}, and dislocation dynamics \cite{koslowski2002phase} to name a few.
In addition to fields which describe the microscopic state of a system (e.g., local temperature, stress/strain, etc.) phase field models introduce one or more continuous fields which describe the local phase of the system, implicitly enabling boundary tracking between different mesostates. 
The phase field variables take on distinct values within each phase and vary continuously between these values through a small but finite sized boundary between regions of differing phases. 
The governing equations for both the physical and phase fields are typically defined through a chosen thermodynamic free energy functional of the form \cite{tourret2022}
\begin{equation}\label{eqn:phaseField}
A^{\text{PF}} = \int \diff^n r \ a(c_1,\ldots,c_k,\Phi_1,\ldots,\Phi_l,\nabla c_1,\ldots,\nabla c_k,\nabla \Phi_1,\ldots,\nabla \Phi_l,T,\varepsilon,\ldots) 
\end{equation}
including both conserved, $c_1,\ldots,c_k$, and and non-conserved, $\Phi_1,\ldots,\Phi_l$, phase field variables, their gradients, and the relevant thermodynamic fields for the problem at hand (e.g., temperature $T$, strain $\varepsilon$, $\ldots$). The free energy density $a$ typically contains bulk free energies of each phase, a multi-welled function of the phase fields with minima for each phase, and terms involving the gradients of the phase fields which contribute to the energetic cost due to the presence of interfaces between phases. 
The governing equations for conservative phase fields are often chosen to obey a Cahn-Hilliard equation \cite{cahn1958} 
\begin{equation*}
\frac{\pd c_i}{\pd t} = \nabla \cdot M^c_{ij}\  \nabla\frac{\delta A^{\text{PF}}}{\delta c_j}
\end{equation*}
whereas non-conservative fields are typically set to obey an Allen-Cahn equation \cite{allen1979}
\begin{equation}\label{eqn:allenCahn}
\frac{\pd \Phi_i}{\pd t} = -M^{\Phi}_{ij} \frac{\delta A^{\text{PF}}}{\delta \Phi_j}.
\end{equation}
Here, $M^c_{ij}$ and $M^{\Phi}_{ij}$ are mobility matrices coupling the velocities of the phase field variables to their conjugate forces. Throughout, we shall make use of the Einstein summation convention, summing all repeated indices, but we will not distinguish between raised and lowered indices.   
Although thermodynamic consistency would allow for coupling terms between these two sets of equations \cite{reina2015entropy}, such couplings are rarely considered.  
\par
The functional forms for interfacial energies and the multi-well phase field potential are generally set phenomenologically, and both contain parameters which, along with the mobilities, must be tuned to the individual problem at hand. Typically, this is achieved by comparing to experimental data or lower scale computational models. A notable exception is that of lattice models obeying a microscopic master equation, for which Cahn-Hilliard phase field equations (with stochastic noise) have been derived via coarse-graining under suitable assumptions \cite{bronchart2008}. In this case, the phase field parameters are given as explicit exponential averages of interaction energies between neighboring atoms. 
To the best of the authors' knowledge, however, a derivation of non-conservative phase field models in elastic media is still lacking. In this work, we demonstrate how the recently developed Stochastic Thermodynamics with Internal Variables (STIV) framework \cite{stiv}, rooted in non-equilibrium statistical mechanics, can be leveraged to produce an Allen-Cahn type phase field model without parameter fitting, including both the functional free energy and mobility coefficients, of a system with a traveling front. 
\par
The \stiv framework generates ab initio non-equilibrium thermodynamics models directly from microscopic physics. 
Given a microscopic system which is well described by Langevin dynamics and given a parametric family of probability densities to approximate the density of states, the \stiv framework produces a thermodynamic model with internal variables \cite{maugin1994} with closed form equations for the evolution of the internal variables (the parameters of the density of states) and explicit equations for non-equilibrium quantities including work rate, heat rate, and the rate of total entropy production. When the approximate density is taken to be a Gaussian distribution, the resulting equations obey a gradient flow with respect to the non-equilibrium free energy, which is a particular case of the so-called GENERIC structure \cite{grmela1997}. 
There is no need to posit a phenomenological form for the non-equilibrium free energy nor for the evolution equations for the internal variables. They are supplied directly by the framework in accordance with stochastic thermodynamics \cite{seifert2012} and a non-equilibrium variational principle \cite{eyink1996}, as will be described in detail in the text.
\par
We use the \stiv framework to derive a non-conservative (Allen-Cahn type) phase field model from microscopic Langevin equations in a one dimensional colloidal system exhibiting a moving phase front.
The system in question is a one dimensional mass-spring-chain with double-well interaction energies between neighboring masses. 
The \stiv method is capable of numerically solving for arbitrary interactions, but in certain cases closed form equations can be obtained. 
We analyze two interaction potentials analytically (a piecewise quadratic double-well potential and a quartic potential), and evaluate a realistic interaction energy obtained from simulation data numerically. 
Interestingly, the resulting evolution equations are not of the form of Eq.~\eqref{eqn:allenCahn}, but are instead coupled to the equilibrium equations (i.e., the evolution equations for $\mu$). That is, denoting by $\mu$ the particles' mean position and $\Phi$ the probability that a given spring is in the rightmost well (the non-conservative phase field variable), the resulting equations are of the form,
\begin{equation} \label{Eq:General}
\mat{\dot{\mu} \\ \dot{{\Phi}} \\ } = -M \mat{ \frac{\pd \hat{A}^{\noneq}}{\pd \mu} \\ \frac{\pd \hat{A}^{\noneq}}{\pd \Phi } \\},
\end{equation}
where $\hat{A}^{\noneq}$ is the non-equilibrium free energy obtained through \stiv and $M$ is a mobility matrix which is not block diagonal, hence inducing a coupling between the two sets of equations. Remarkably, the expression of $\hat{A}^{\noneq}$ for the two analytical examples considered did not contain a term proportional to  $|\nabla \Phi|^2$, as it is standard in phase field modeling.
The \stiv model provides a number of simplifying features providing an advantage over traditional phase field models.
First, the functional form of the \stiv model is entirely given from statistical mechanics principles and it has no free parameters, i.e., it does not need to be fit to experimental or computational data. Second, it can successfully capture the behavior of the system under arbitrary external pulling protocol. Finally, the \stiv model naturally captures stress-driven nucleation of the propagating phase front for all of the examples tested.  We remark though, that the strong physical fidelity of the \stiv model implies that the interface thickness is realistic in size. This is often avoided in traditional phase field models to alleviate the computational cost of the numerical simulations.
\par
The layout of the paper is as follows; we begin in Section \ref{sec:STIV} with a brief explanation of the \stiv framework intended to assist readers in their own implementation. In Section \ref{sec:results}, we derive the \stiv model for the one dimensional colloidal system with arbitrary nearest neighbor interaction energy which will result in equations of the form given in \eqref{Eq:General}. 
We also compute analytically the non-equilibrium free energy and mobilities for a piecewise quadratic and quartic interaction potentials.
Next, we introduce two numerical techniques in Section \ref{sec:verification} which allow one to evaluate the \stiv model for any interaction potential. 
We provide computational verification of these two numerical methods against analytical evaluation and Langevin simulations. These two numerical methods are then used to produce an \stiv model for phase front propagation in coiled-coil proteins using a realistic double-well free energy gathered from molecular dynamics simulations.
Finally, in Section \ref{sec:discussion}, we provide a few concluding remarks. 

\section{Method: Stochastic Thermodynamics with Internal Variables} \label{sec:STIV}
In this section, we give a brief overview of the \stiv framework intended to assist practitioners in producing their own implementation. 
Those interested in the derivation of the framework are referred to \cite{stiv}.
The \stiv framework is a unification of the stochastic thermodynamics definitions for thermodynamic quantities at the level of microscopic trajectories \cite{seifert2012} with the variational scheme proposed by Eyink for approximating the evolution of probability densities which arise in non-equilibrium settings \cite{eyink1996}.
For a system of interest which is well described by a stochastic Langevin equation at constant temperature, one posits a functional form of a parametric probability distribution $\hat{p}(x,\alpha)$ to serve as an approximation to the true density of states $p(x,t)$. Here, the parameters are labeled $\alpha$ and correspond to the internal variables of the macroscopic thermodynamic model. 
The variational scheme is applied to the parametric density to produce ordinary differential equations governing the dynamics of the parameters $\alpha(t)$ so that the evolution of the parametric density extremizes the weak form of the true evolution (given by the Fokker-Planck equation) at all times. 
In particular, those differential equations are 
\begin{equation}\label{Eq:EyinkEquations}
\avg{ \frac{\pd\log(\hat{p}) }{\pd \alpha_i} \ \frac{\pd\log(\hat{p}) }{\pd \alpha_j}}_{\hat{p}} \dot{\alpha}_j = \avg{ \fL^\dagger \frac{\pd\log(\hat{p}) }{\pd \alpha_i} }_{\hat{p}},
\end{equation}
where $\fL^\dagger$ is the adjoint to the Fokker-Plank operator associated with the Langevin dynamics\footnote{Given the Langevin dynamic equation $\diff x = -\frac1{\eta}\frac{\pd}{\pd x} e(x,\lambda)\, \diff t + \sqrt{2d}\, \diff w$, the corresponding Fokker-Planck equation is given by $\frac{\pd p}{\pd t} = \fL p \equiv \frac1{\eta}\frac{\pd}{\pd x}(\frac{\pd e}{\pd x}\ p) + d\  \frac{\pd^2 p}{\pd x^2}$ and $\fL^\dagger$ reads $\fL^\dagger \psi = -\frac{1}{\eta} \frac{\pd e}{\pd x} \frac{\pd \psi}{\pd x} + d \frac{\pd^2 \psi}{\pd {x^2}}$. } and $\avg{ }_{\hat{p}} $ denotes the ensemble average with respect to the approximate distribution $\hat{p}$, i.e., $\avg{ g(x,\alpha) }_{\hat{p}} =\int g(x,\alpha) \hat{p}(x,\alpha) \, dx $. In a simple system and when $\hat{p}(x,\alpha)$ has a simple form, expectations can be computed analytically, but otherwise these must be approximated via other means.
The parametric density is then substituted for the true density of states in the definitions of macroscopic thermodynamic quantities given by stochastic thermodynamics, including those for the work rate, heat rate, and rate of total entropy production. 
Assuming the system is governed by a conservative potential energy $e(x,\lambda)$ where $\lambda$ is an external control protocol, the following equations give the approximation to the relevant thermodynamic quantities: the approximate energy is $\hat{E}(\lambda,\alpha) = \avg{e}_{\hat{p}}$, the approximate entropy is $\hat{S}(\alpha) = \avg{-k_B \log(\hat{p})}_{\hat{p}}$ where $k_B$ is the Boltzmann constant, the approximate non-equilibrium free energy is $\hat{A}^{\noneq}(\lambda,\alpha) = \hat{E} - T\hat{S}$ where $T$ is the absolute temperature, the work rate is $\frac{\diff}{\diff t}\hat{W}(\lambda,\alpha) = \frac{\pd}{\pd\lambda} \hat{A}^{\noneq} \dot{\lambda}$, the rate of heat flowing into the system is $\frac{\diff}{\diff t}\hat{Q}(\lambda,\alpha) = \frac{\pd \hat{E}}{\pd \alpha_i}\ \dot{\alpha}_i$, and the rate of total entropy production is $\frac{\diff}{\diff t} T\hat{S}^{\tot}(\lambda,\alpha) = - \frac{\pd \hat{A}^{\noneq}}{\pd\alpha_i}\  \dot{\alpha}_i$. 
The resulting structure of these thermodynamic equations exactly matches those of the commonly used thermodynamics with internal variables formalism \cite{maugin1994}, with the parameters of the parametric probability distribution, $\alpha$, serving as the internal variables. 
For specific choices of the approximate density, Eqs.~\eqref{Eq:EyinkEquations} can be shown to be equivalent to gradient flow dynamics with respect to the non-equilibrium free energy, ensuring the non-negativity of the entropy production. 
This is the case for the multivariate Gaussian density which is used in what follows. 
\section{Results} \label{sec:results}
As mentioned above, we use the \stiv framework to derive microscopically-based kinetic equations for the phase fraction for a colloidal mass-spring-chain system with double-well interactions. We assume that the total energy is given as the sum of pairwise interactions between neighbors, and that the pairwise interaction, $u(r)$, is double-welled with $r=0$ defining the boundary point between wells (for simplicity). For a system with $N$ free masses with positions $x = (x_1,\ldots,x_N)$ attached in series with one end held fixed, and the other attached to an external control with position $\lambda$, the total energy of the system is given by 
\begin{equation*}
e(x,\lambda) = u(x_1) + \sum_{i=2}^N u(x_i - x_{i-1}) + u(\lambda - x_N).
\end{equation*}
Assuming a time dependent external protocol, $\lambda(t)$, the governing stochastic Langevin equation for the mass positions is 
\begin{equation}
\diff x = -\frac1{\eta}\frac{\pd e}{\pd x} (x,\lambda)\, \diff t + \sqrt{2d}\, \diff w \label{eqn:Langevin}
\end{equation}
where $\eta$ is the drag coefficient, $d = 1/\eta\beta$ is the diffusion coefficient, $\beta = 1/k_B T$ is the inverse absolute temperature in energy units, and $w = (w_1,\ldots,w_N)$ is a vector of independent standard Brownian motions \cite{risken1989fokker}. The true density of states associated with the mass positions obeys a Fokker-Planck equation given by 
\begin{equation*}
\frac{\pd p}{\pd t} = \fL p \equiv \frac1{\eta}\frac{\pd}{\pd x}\left(\frac{\pd e}{\pd x}\ p\right) + d\  \frac{\pd^2 p}{\pd x^2}. 
\end{equation*}
Since, in many instances, one cannot solve this equation outright for $p$, we choose an approximate parameterized density $\hat{p}(x,\alpha(t))$, and use the \stiv formalism to determine the kinetic equations for the internal variables $\alpha$, as well as approximations to any mechanical or thermodynamic quantity of interest. 
\par 
Previously in \cite{stiv}, the \stiv framework was used to study this same system in the particular case that the interaction potential was piecewise quadratic with a cusp at the origin. There, $\hat{p}$ was assumed to be a multivariate Gaussian, with internal variables given by the mean $\mu$ and the covariance matrix $\Sigma$ of the mass positions. The resulting kinetic equations for $\mu$ and $\Sigma$ were shown to be a gradient flow with respect to the non-equilibrium free energy. Moreover, the non-equilibrium free energy implicitly depended on the approximate probability that a given spring will fall in the right (positive) well, 
\begin{equation*}
\hat{\Phi}_i \equiv \avgph{\Indic_{(x_i - x_{i-1} > 0)} }
\end{equation*}
much like in traditional phase field models (with a few notable difference which are mentioned below).
This leads one to ask whether it is possible to change variables to promote $\hat{\Phi}$ from a dependent variable (namely a function of $\mu$ and $\Sigma$) to an independent variable. 
\par
To answer this question we will in general let $u$ be any interaction potential, and take for the approximate density of states, $\hat{p}(x,\alpha)$ a strictly diagonal multivariate Gaussian, with internal variables corresponding to the mean mass positions, $\mu = (\mu_1,\ldots,\mu_N)$, and standard deviation of the mass positions $\sigma = (\sigma_1,\ldots,\sigma_N)$.
The reason for taking only the diagonal, and not the full covariance matrix will be made clear later.  Mathematically, we approximate the density of states with
\begin{equation*}
\hat{p}(x,\alpha)=\hat{p}(x,\mu,\sigma) = \frac1{(2\pi)^{N/2}} \exp(-\frac1{2}\sum_{i=1}^N \frac{(x_i - \mu_i)^2}{\sigma_i^2} - \frac1{2}\sum_{i=1}^N \log(\sigma_i^2) ).
\end{equation*}
Under this assumption, the approximate phase fraction becomes an explicit function of $\mu$ and $\sigma$,
\begin{equation*}
\hat{\Phi}_i = \Phi\left(\frac{\mu_i - \mu_{i-1}}{\sqrt{\sigma_i^2 + \sigma_{i-1}^2}}\right) \qquad i=1,\ldots,N
\end{equation*}
where $\Phi$ is the cumulative distribution function (CDF) of a standard Gaussian (we set $\mu_0 = \sigma_0 \equiv 0$).\footnote{This follows from the fact that if $X$ and $Y$ are independent Gaussian random variables with means $\mu_x$ and $\mu_y$ and variances $\sigma_x^2$ and $\sigma_y^2$ respectively, then $X - Y$ is a Gaussian random variables with mean $\mu_x - \mu_y$ and variance $\sigma_x^2 + \sigma_y^2$. Furthermore, $\avg{\Indic_{(X > 0)}} = \int_0^\infty \phi\left(\frac{x - \mu_x}{\sigma_x}\right)\frac{\diff x}{\sigma_x} = 1 - \Phi\left(\frac{-\mu_x}{\sigma_x}\right) = \Phi\left(\frac{\mu_x}{\sigma_x}\right)$.}
Using the \stiv method with the parameterization above yields the following kinetic equations for the internal variables (see \ref{App:Derivation_KineticEquations} for the full derivation)
\begin{align*}
\dot{\mu}_i &= \frac1{\eta}\left(\hat{f}(\epsilon_i,\tau_i) - \hat{f}(\epsilon_{i+1},\tau_{i+1}) \right) \\ 
 \dot{\sigma}_i&= \frac{\sigma_i}{\eta}\left( \frac{\pd \hat{f}}{\pd\epsilon}(\epsilon_{i+1},\tau_{i+1}) + \frac{\pd \hat{f}}{\pd\epsilon}(\epsilon_i,\tau_i)  + \frac1{\beta\sigma_i^2} \right)
\end{align*}
where 
\begin{equation}\label{eqn:fHat}
\hat{f}(\epsilon,\tau) = \avg{-u'(z)}_{z\sim \N(\epsilon,\tau^2)}
\end{equation}
is the average of the interaction force of a spring that is normally distributed with mean $\epsilon$ and standard deviation $\tau$. Here, we've set $\epsilon_i = \mu_i - \mu_{i-1}$ (with $\mu_0 = 0$, $\mu_{N+1} = \lambda$) to be the mean length of spring $i=1,\ldots,N+1$, and $\tau_i^2 = \sigma_i^2 + \sigma_{i-1}^2$  (with $\sigma_0^2 = \sigma_{N+1}^2 = 0$) to be the variance of spring $i=1,\ldots,N+1$.
For convenience, we shall write $\hat{f}_i \equiv \hat{f}(\epsilon_i,\tau_i)$.
Since we have utilized a multivariate Gaussian approximation, we are guaranteed that the kinetic equations obey a gradient flow with respect to the non-equilibrium free energy (see \cite{stiv} and \ref{App:Derivation_KineticEquations})
\begin{equation*}
\eta\dot{\mu} = - \frac{\pd \hat{A}^{\noneq}}{\pd \mu} \qquad
\eta\dot{\sigma} = -\frac{\pd \hat{A}^{\noneq}}{\pd \sigma} 
\end{equation*}
with $\hat{A}^{\noneq} \equiv \hat{E} - T\hat{S} = \avg{e}_{\hat{p}} + k_BT \avg{\log(\hat{p})}_{\hat{p}}$. 
For a general interaction potential the approximate non-equilibrium free energy is 
\begin{equation}\label{eqn:neqFE}
\hat{A}^{\noneq} = \sum_{i=1}^{N+1} \hat{U}(\epsilon_i,\tau_i) - \frac1{2\beta}\sum_{i=1}^N \log(\sigma_i^2) - \frac{N}{2\beta}\left(1 + \log(2\pi)\right)
\end{equation}
where 
\begin{equation*}
    \hat{U}(\epsilon,\tau) = \avg{u(z)}_{z \sim \N(\epsilon,\tau^2)}
\end{equation*}
is the average interaction energy of a spring whose length is normally distributed with mean $\epsilon$ and standard deviation $\tau$.
The work rate, heat rate, and rate of total entropy production must take the form (see \ref{App:ThermodynamicsEquations})
\begin{align*}
\frac{\diff}{\diff t} \hat{W} &=  \frac{\pd \hat{A}^{\noneq}}{\pd\lambda} \dot{\lambda} = -\hat{f}_{N+1}\dot{\lambda} \\
\frac{\diff}{\diff t} \hat{Q} &= \frac{\pd \hat{E}}{\pd \alpha} \dot{\alpha} =  -\eta \abs{\dot{\mu}}^2 - \eta \abs{\dot{\sigma}}^2 + \frac1{\beta}\sum_{i=1}^N \frac{\dot{\sigma}_i}{\sigma_i}\\ 
T \frac{\diff }{\diff t} \hat{S}^{\tot} &= -\frac{\pd \hat{A}^{\noneq}}{\pd \alpha} \dot{\alpha} = \eta \abs{\dot{\mu}}^2 + \eta \abs{\dot{\sigma}}^2.
\end{align*}
Returning to the non-equilibrium free energy, we highlight that in \cite{stiv} the piecewise quadratic double-well interaction potential of the form 
\begin{equation*}
    u(r) = \begin{cases} \frac{k_1}{2}(r + l_1)^2 & r < 0 \\ \frac{k_2}{2}(r - l_2)^2 + h_2 & r \geq 0 \\ \end{cases},
\end{equation*}
where $l_1$ and $l_2$ are the positive distances to the two quadratic energy minima, $k_1$ and $k_2$ are the two stiffnesses, and $h_2 = k_1l_1^2/2 - k_2l_2^2/2$ is chosen to ensure continuity, yielded (see Section 5 of the Supplemental Material for \cite{stiv})
\begin{align}\label{eqn:energyBiquadratic}
 \hat{U}(\epsilon,\tau)  &= \frac{k_1}{2}(\tau^2 + (\epsilon + l_1)^2)\left(1 - \Phi(\epsilon/\tau)\right)  \\
    &\quad + \left(\frac{k_2}{2}(\tau^2 + (\epsilon-l_2)^2) + h_2\right)\Phi(\epsilon/\tau)  \nonumber\\
    &\quad + \left(\frac{k_2 - k_1}{2}\epsilon - (l_1k_1 + l_2k_2)\right)\tau\phi(\epsilon/\tau), \nonumber
\end{align}
where $\phi$ is the probability density function (PDF) of a standard Gaussian. This approximate interaction energy decomposes as an energetic contribution from the left well multiplied by $(1 - \Phi)$, the right well contribution multiplied by $\Phi$, along with a interfacial term multiplied by $\phi$ which tends to zero rapidly when $\abs{\epsilon/\tau} >> 1$ (recall that the boundary between wells is chosen to be $\epsilon = 0$). 
As mentioned above, for a multivariate Gaussian $\hat{\Phi}_i = \int \Indic_{(x_i - x_{i-1} > 0)} \hat{p}(x,\mu,\sigma)\diff x = \Phi\left(\frac{\epsilon_i}{\tau_i}\right)$. 
Since $\hat\Phi_i$ is a local function of $\tau_i$, after the change of variables presented in the following section $\tau_i$ becomes a local function of $\hat\Phi_i$. Thus, we see that the internal energy of each spring will always be a local function of the phase variable. The entropy (the remaining portion of Eq. \ref{eqn:neqFE}), on the other hand, gives rise to a non-local function of $\hat\Phi$ which significantly differs from the $|\nabla\Phi|^2$ term found in the Landau-Ginzburg free energy.  Moerover, the STIV free energy lacks any explicit dependence on $\nabla \Phi$.  In the following section, we qualitatively compare plots of the STIV free energy with that of a traditional phase field model.
\par
The explicit appearance of the phase fraction in the approximate interaction energy, $\hat{U}$, and hence the non-equilibrium free energy, $\hat{A}^{\noneq}$, is not guaranteed for all interaction potentials. For example, in the case of a double-well quartic interaction energy with minima at $-l_1$ and $l_2$, 
\begin{equation}\label{eqn:quartic}
u(r) = \frac{r^4}{4} + (l_1 - l_2)\frac{r^3}{3} - l_1l_2\frac{r^2}{2}, 
\end{equation}
the average interaction energy of each spring reads
\begin{equation*}
\hat{U}(\epsilon,\tau) = \frac{\epsilon^4}{4} + (l_1 - l_2)\frac{\epsilon^3}{3} - l_1l_2\frac{\epsilon^2}{2} + \tau^2\left(\frac{3}{2}\epsilon^2 + (l_1 - l_2)\epsilon -\frac{l_1l_2}{2}\right) + \frac{3}{4}\tau^4,
\end{equation*}
which contains no explicit dependence on the phase fraction. Hence it will not appear explicitly in the kinetic equations for $\mu$ or $\sigma$. Despite this, it is always possible to promote the phase fraction $\hat{\Phi}$ to an independent variable, producing a phase field model from STIV, as will be shown in the next section.
\par

\subsection{Change of variables}
In the previous section, we showed how to apply the STIV framework to a one dimensional mass-spring-chain with arbitrary interaction energies. However, the phase fraction $\hat{\Phi}$ was a dependent function of the internal variables, $\mu$ and $\sigma$, and it was therefore unclear whether the resulting evolution equations could be written as a function of $\mu$ and $\Phi$ only, as it is typical in phase field equations. In this section we demonstrate how to construct a mapping $(\mu,\sigma) \rightarrow (\mu,\hat{\Phi})$ to promote the phase fraction to an independent variable. This is the reason for reducing the degrees of freedom of the internal variable approximation from a full covariance matrix, used in \cite{stiv}, to just the diagonal in the analysis presented here, (i.e. $\Sigma_{ij} \rightarrow \delta_{ij}\sigma_i^2$). Since the full covariance consists of $N(N-1)/2$ independent variables whereas $\hat{\Phi}$ is an $N$ dimensional vector ($\hat{\Phi}_{N+1}$ can be written as a function of the other terms and $\lambda$), a map from the full covariance matrix to the phase fractions could never be invertible. However, the mapping $\hat{\Phi} = \left(\Phi(\epsilon_1/\tau_1),\ldots,\Phi(\epsilon_N/\tau_N)\right) \rightarrow \sigma$ is one-to-one away from the transition points, $\epsilon_i  = \mu_i - \mu_{i-1} \neq 0$ for all $i \in 1,\ldots,N$. This is because the cumulative distribution function of a standard Gaussian, $\Phi$, is an increasing function, and hence invertible, and one can invert the relationship between $\sigma$ and $\tau$ through 
\begin{equation*}
\sigma_i = \sqrt{\sum_{j=0}^{i-1} (-1)^j\tau_{i-j}^2 }.
\end{equation*}

\par 
Let $\Phi^{-1}$ be the inverse cumulative distribution function of a standard Gaussian, and for $\hat{\Phi}_j \neq 1/2$ for all $j$, let 
\begin{align*}
\tau_i(\mu,\hat{\Phi}) &= \frac{\mu_i - \mu_{i-1}}{\Phi^{-1}(\hat{\Phi}_i)} \\ 
\sigma_i(\mu,\hat{\Phi}) &= \sqrt{\sum_{j=0}^{i-1}(-1)^j \tau_{i-j}(\mu,\hat{\Phi})^2 } \\
\hat{\phi_i}(\mu,\hat{\Phi}) &= \phi\circ\Phi^{-1}(\hat{\Phi}_i). 
\end{align*}
We remark that this change of variables is one-to-one in the one-dimensional mass-spring system considered, where the number of free masses, and hence, their associated standard deviations $\sigma$, is equal to the number of independent spring lengths and associated phase fractions $\hat{\Phi}$, thought this may not always be the case for mass-spring systems in higher dimensions.

Returning to the governing equations of STIV,  a straightforward application of the chain rule to Eq.~\eqref{Eq:EyinkEquations} verifies that under an abstract change of variables $\alpha' = \alpha'(\alpha)$, the velocity of the internal variables changes like a typical velocity vector, i.e., $\dot{\alpha'}_i = \frac{\pd \alpha'_i}{\pd \alpha_j}\dot{\alpha}_j$.
Using this fact and the above definitions, we can use the implicit formula for $\hat{\Phi}$ to determine the equivalent dynamical equations for $\mu$ and $\hat{\Phi}$ as those for $\mu$ and $\sigma$. The equations for $\dot{\mu}$ remains unchanged while 
\begin{align*}
    \frac{\diff}{\diff t} \hat{\Phi}_i &= \hat{\phi_i}\left(\frac{\dot{\mu}_i - \dot{\mu}_{i-1}}{\tau_i} - \frac{(\mu_i - \mu_{i-1}) \dot{\tau}_i}{\tau_i^2} \right) \\  
    &= -\frac{\hat{\phi}_i}{\eta} \left(\frac{\Delta \hat{f}_i }{\tau_i} + \frac{(\mu_i-\mu_{i-1})}{\tau_i^3}\left(\sigma_i^2\left(\frac{\pd \hat{f}_{i+1}}{\pd \epsilon} + \frac{\pd\hat{f}_i}{\pd \epsilon}\right) + \sigma_{i-1}^2\left(\frac{\pd\hat{f}_i}{\pd \epsilon} + \frac{\pd\hat{f}_{i-1}}{\pd \epsilon}\right) + \frac{2}{\beta} \right)\right),
\end{align*}
where $\Delta \hat{f}_i = \hat{f}_{i+1} - 2\hat{f}_i + \hat{f}_{i-1}$ is the discrete Laplacian of $\hat{f}$. 
Although these equations are only well defined away from phase transition points, they are equivalent to the previous dynamical equations for $\mu$ and $\sigma$ which produce smoothly varying functions of time, and hence we find that the new equations can be numerically integrated through phase transition points without loss of accuracy so long as the integration step size is small and the integration scheme never exactly evaluates at $\hat{\Phi}_i = 1/2$. This is because at transitions, both $\epsilon_i = \mu_i - \mu_{i-1}$ and $\Phi^{-1}(\hat{\Phi}_i)$ tend to zero so that 
$\tau_i = (\mu_i - \mu_{i-1})/\Phi^{-1}(\hat{\Phi}_i)$ is finite, leaving $\tau_i$ continuous in time.
However, this makes it challenging (or even impossible) to numerically integrate the equations for $\mu$ and $\hat{\Phi}$ when starting at or near a transition.
\par
We highlight the effect of the change of variables on a non-equilibrium free energy density, Eq.~\eqref{eqn:neqFE}, in the case of a single free mass (and hence no non-local contributions) in the quartic potential and no external control. In panel (A), we see $\hat{A}^\noneq$ as a function of $\epsilon$ and $\tau$. Larger values of the free energy are shown in yellow, while smaller values are shown in dark blue. For small values of $\tau$, $\hat{A}^\noneq$ is quartic with a double-well in $\epsilon$. But as $\tau$ increases, the minima coalesce to form a single-well potential. The minimum free energy transition pathway between the two minima is highlighted by the dotted yellow line. As one would expect, during a transition the variance in the strain is increased because the true density of states is multi-modal (this phenomena can be observed in \cite{stiv} Movie S1). $\hat{A}^\noneq$ is shown as a function of $\epsilon$ and $\hat{\Phi}$ in panel (B). The regions in white correspond regions where $\hat{A}^\noneq = \infty$, due to the restriction that $\tau = \epsilon/\Phi^{-1}(\hat{\Phi}) \geq 0$. This amounts to a strong enforcement of $\epsilon < 0$ when $\hat{\Phi} < 1/2$ and $\epsilon > 0$ when $\hat{\Phi} > 1/2$. In a traditional phase field free energy (again, without non-local contributions), like the one shown in panel (C), this constraint is not imposed but the choice of bulk free energies and double-well function in the phase field variable generally make this configuration energetically favorable (here we choose $\hat{\Phi}(1 - \hat{\Phi})$ for the double-well function as in \cite{apel2006}). The yellow dashed lines in both panel (B) and (C), highlighting the minimum free energy transition paths, reveal that both models predict qualitatively similar transition paths.
\begin{figure}
    \centering
    \includegraphics[width=\linewidth]{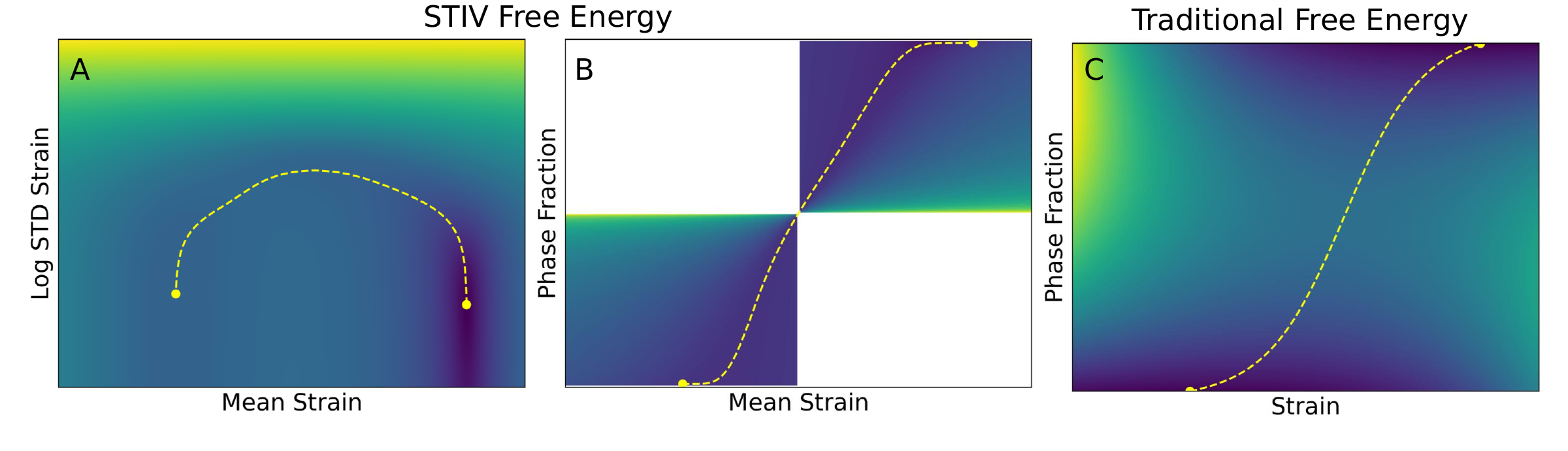}
    \caption{\label{fig:FE}A comparison of the \stiv non-equilibrium free energy for a single particle in a quartic potential with no external control to a free energy density more similar to one used in traditional phase field (non-local terms from both free energies are neglected here). Larger values are shown in light yellow, and smaller values are shown in dark blue. In panel (A), we see $\hat{A}^\noneq$ as a function of $\epsilon$ and $\tau$ before the change of variables. In panel (B), $\hat{A}^\noneq$ is shown as a function of $\epsilon$ and $\hat{\Phi}$. Regions in white correspond to $\hat{A}^\noneq = \infty$. Panel (C) shows a more traditional phase field free energy as a function of $\epsilon$ and $\hat{\Phi}$. Here, $\hat{A}^\noneq(\epsilon,\tau)$, $\tau$ small and fixed,  is approximated at the left and right minima with single-well quartic (i.e., no odd order terms) functions of $\epsilon$ to serve as the phase field free energies of the two phases. The two minima are then interpolated using  linear interpolation (that is, $(1-\hat\Phi)$ and $\hat\Phi$) and the double-well function $c\,\hat\Phi(1-\hat\Phi)$  with large coupling strength $c$ is included to highlight the impact of the double-well on the total free energy. All three panels show the minimum free energy transition pathway between left and right minima as a yellow dashed line.}
\end{figure}
\par 
Finally, it is worth noting that this change of variables can be defined for an arbitrary interaction potential. Under this change of variables, the system is fully described by $\mu$ and the phase fractions $\hat{\Phi}$, turning the original \stiv model into a phase field model. We show next that the phase field model still obeys a gradient flow dynamics with respect to the non-equilibrium free energy.
\par 
It turns out that we need not re-derive the thermodynamic equations under the change of variables as all three are invariant under the transformation. For example, consider the rate of total entropy production, and the abstract change of variables $\alpha' = \alpha'(\alpha)$. We see that
\begin{equation*}
T\frac{\diff}{\diff t}\hat{S}^{\tot} = -\sum_{i=1}^N \frac{\pd \hat{A}^{\noneq}}{\pd \alpha_i} \dot{\alpha}_i = -\sum_{i,j,k}^N \frac{\pd \alpha'_j}{\pd \alpha_i} \frac{\pd \hat{A}^{\noneq}}{\pd \alpha'_j} \frac{\pd \alpha_i}{\pd \alpha'_k} \dot{\alpha}'_k = -\sum_{j,k}^N \delta_{jk} \frac{\pd \hat{A}^{\noneq}}{\pd \alpha'_j}\dot{\alpha}'_k = -\sum_{j=1}^N \frac{\pd \hat{A}^{\noneq}}{\pd \alpha_j'} \dot{\alpha}_j'. 
\end{equation*}
It is clear that this holds for both the work rate and the heat rate as well. What does change, however, is the structure of the gradient flow. In particular, under $\mu$ and $\sigma$, the dissipation matrix $M$ was diagonal, i.e.,
\begin{equation}
    \dot{\alpha}_i=-M_{ij} \frac{\partial \hat{A}^{\noneq}}{\partial \alpha_j}=-\frac{1}{\eta}\frac{\partial \hat{A}^{\noneq}}{\partial \alpha_i}.
\end{equation}
Under the change of variables, the dissipative matrix transforms like a contravariant two tensor 
\begin{equation*}
M_{ij}' = \frac{\pd \alpha'_i}{\pd \alpha_k}\frac{\pd \alpha'_j}{\pd \alpha_l} M_{kl}.
\end{equation*}
This means the new gradient flow structure is given by 
\begin{equation*}
\mat{\dot{\mu} \\ \dot{\hat{\Phi}} \\ } = -\frac{1}{\eta} \mat{ \text{Id} & \left(\frac{\pd \hat{\Phi}}{\pd \mu}\right)^T \\  \frac{\pd \hat{\Phi}}{\pd \mu} & \frac{\pd \hat{\Phi}}{\pd \mu}\left(\frac{\pd \hat{\Phi}}{\pd \mu}\right)^T + \frac{\pd \hat{\Phi}}{\pd \sigma}\left(\frac{\pd \hat{\Phi}}{\pd \sigma}\right)^T \\ } \mat{ \frac{\pd \hat{A}^{\noneq}}{\pd \mu} \\ \frac{\pd \hat{A}^{\noneq}}{\pd \Phi } \\}. 
\end{equation*}
Note that although the equation for $\dot{\mu}$ remains the same for both sets of variables, $\frac{\pd \hat{A}^{\noneq}}{\pd \mu}\Big|_{\sigma \text{ fixed}} \neq\frac{\pd \hat{A}^{\noneq}}{\pd \mu}\Big|_{\hat{\Phi} \text{ fixed}}  $. Thus,  $\dot{\mu}$ now depends on $\frac{\pd \hat{A}^{\noneq}}{\pd \hat{\Phi}}$ after the change of variables, and similarly, $\dot{\hat{\Phi}}$ also depends on $\frac{\pd \hat{A}^{\noneq}}{\pd \mu}$. That is, the dynamics of $\mu$ and $\hat\Phi$ in the STIV phase field model become coupled after the change of variables. This represents another significant difference with traditional phase field models, where, typically, the dynamics are assumed to be decoupled, i.e., 
\begin{equation*}
\mat{\dot{\mu} \\ \dot{\hat{\Phi}} \\} = -\mat{M^\mu & 0 \\ 0 & M^\Phi} \mat{ \frac{\pd A^{\text{PF}}}{\pd \mu} \\ \frac{\pd A^{\text{PF}}}{\pd \Phi} }.
\end{equation*}
\par
Thus, we see that through a change of variables, the phase field model derived through the \stiv method is a gradient flow of the non-equilibrium free energy.  Of particular importance is the fact that the non-equilibrium free energy and the dissipative matrix produced by the \stiv method can be computed analytically in simple cases or numerically when an analytical solution is impossible. In the following section we will demonstrate how the \stiv method can be implemented numerically to allow for arbitrary interaction potentials. 

\section{Numerical methods} \label{sec:verification}
In this section, we introduce two methods for evaluating \stiv models numerically. The ability to implement \stiv numerically is a necessity, as only a few simple systems offer exact solutions. In particular, one would like to be able to utilize interaction potentials gathered from physical or computational experiments to produce realistic models and predictions. 
\par 
We propose two methods of numerical implementation, one specific to the multivariate Gaussian approximation to the density of states based on Gauss-Hermite quadrature, and another more general method which makes use of random sampling. The benefits of the Gauss-Hermite method are its accuracy and fast run time when compared to the sampling method, which is generally less accurate for a fixed computational time. Both methods provide approximations to $\hat{f}(\epsilon,\tau)$ in Eq.~\eqref{eqn:fHat} to be used in the dynamical equations for $\mu$ and $\sigma$ (or equivalently $\mu$ and $\hat{\Phi}$). As the name suggests, the Gauss-Hermite method numerically approximates $\hat{f}(\epsilon,\tau)$ using a Gauss-Hermite quadrature rule \cite{abramowitz1948} every time $\dot{\mu}$ and $\dot{\sigma}$ (or $\dot{\hat{\Phi}}$) are computed, with $\pd_\epsilon \hat{f}(\epsilon,\tau)$ computed using a two-point finite difference method. 
\par
Alternatively, the sampling method precomputes a polynomial approximation to $\hat{f}(\epsilon,\tau)$ using independent and identically distributed (i.i.d.) random samples on a user defined grid. In this implementation, Chebyshev points of order $N_\epsilon$ and $N_\tau$ are used on a box $B \subset \mathbb{R}\times \mathbb{R}_{\geq0}$. At each grid point $\{(e_i,t_j)\in B \mid i\in 1,\ldots, N_\epsilon,\ j \in 1,\ldots, N_\tau\}$, $\hat{f}(e_i,t_j)$ is approximated using the empirical average of $N_s$ samples from a Gaussian distribution with mean $e_i$ and variance $t_j^2$. Mathematically, we denote the empirical average at the sample points by
\begin{equation*}
  \tilde{f}(e_i,t_j) \equiv \frac1{N_s}\sum_{k=1}^{N_s} -u'(Z_k) \qquad \{Z_k\}_{k=1}^{N_s} \sim \N(e_i,t_j^2)\ \text{  i.i.d.}
\end{equation*}
$\hat{f}(\epsilon,\tau)$ is finally determined as the least squares best approximating polynomial of order $P_\epsilon$ in $\epsilon$ and $P_\tau$ in $\tau$ evaluated using the sample points. Thus, 
\begin{equation*}
\hat{f}(\epsilon,\tau) \equiv \sum_{k,l=0}^{P_\epsilon,P_\tau} c_{kl}\,\epsilon^k\tau^l
\end{equation*}
where the coefficients $c_{kl}$ are chosen to minimize
\begin{equation*}
  L(c) = \sum_{i,j = 1}^{N_\epsilon,N_\tau} \left(\tilde{f}(e_i,t_j) - \sum_{k,l=0}^{P_\mu,P_\sigma} c_{kl}\,e_i^kt_j^l \right)^2.
\end{equation*}
$\frac{\pd\hat{f}}{\pd \epsilon}$ is approximated through polynomial differentiation
\begin{equation*}
\frac{\pd\hat{f}}{\pd \epsilon}(\epsilon,\tau) = \sum_{k=1,l=0}^{P_\epsilon,P_\tau} c_{kl}\,k\epsilon^{k-1}\tau^l.
\end{equation*}

\begin{figure}[h!]
  \begin{center}
    \includegraphics[width = .95\textwidth]{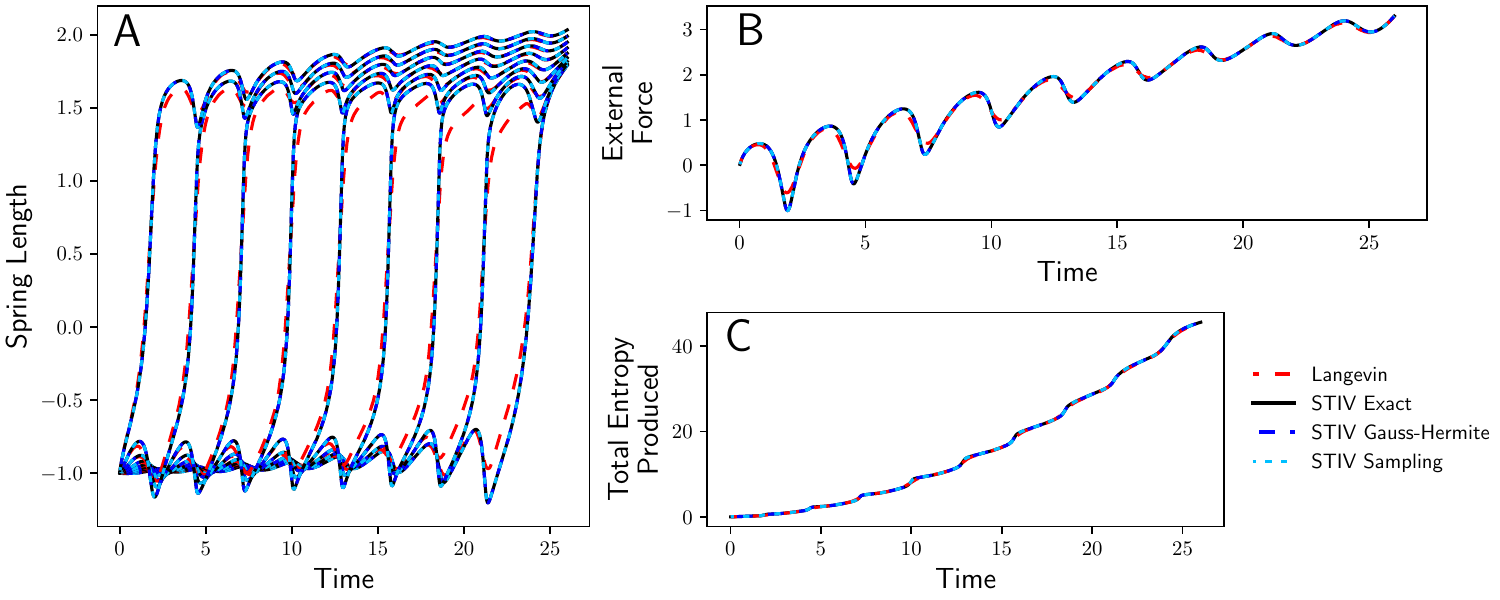}
  \caption{\label{fig:4res} A comparison of the average spring length (A), average external force (B), and total entropy produced (C) for the mass-spring-chain system with quartic interaction potential and linear external pulling protocol. The results are shown for Langevin simulations (red dashes), exact STIV (black solid), Gauss-Hermite STIV (blue dashed), and sample based STIV (light blue dotted).}
  \end{center}
\end{figure}

In Fig.~\ref{fig:4res}, we show a comparison of the \stiv model, including the analytical result, the Gauss-Hermite approximation, and sampling approximation against Langevin simulations for the mass-spring chain with quartic interaction potential and a linear external pulling protocol. Panel (A) shows the average spring lengths, panel (B) shows the average external force, and panel (C) shows the total entropy produced as a function of time.
In every case, we see that the analytical, Gauss-Hermite, and sampling versions of the \stiv framework all overlap, highlighting the accuracy of the STIV model as well as the accuracy of the two numerical implementations. In panel (A), we see the \stiv model does well at capturing the average length of each spring and thus the phase transitions of each spring crossing from left to right well. 
In particular, we see that the \stiv model resolves the phase front interface width at the physical scale. This should be contrasted with typical phase field simulations which often artificially increase the width of the interface to reduce simulation run time. 
\par
Panels (B) and (C) reveal one of the primary benefits of the \stiv model. Since the \stiv model is grounded in physics, it provides quantitative estimates to the external force and total entropy production. We emphasize that the STIV model was directly obtained from the knowledge of the microscopic dynamics and it does not contain any fitting parameters whatsoever. Moreover, the \stiv models produce rapid calculations of the total entropy produced relative to estimation from Langevin simulations. Ignoring the computational cost of obtaining samples for the Langevin simulations, estimation of the total entropy production using Gaussian kernel density estimation requires over $10^4$ times more computational time than every \stiv implementation\footnote{Time estimates are based on $10,000$ Langevin samples and were run on a standard desktop.} (see \cite{stiv} SI Section 9 for details on estimating the rate of entropy production from Langevin samples). As a final note, we highlight the fact that the \stiv model can nucleate a phase transition due to external driving, which is typically done in traditional phase field models by artificially inserting nuclei at a given rate \cite{simmons2000phase}, or by making the dynamical equations stochastic \cite{plapp2011remarks}; see \cite{clouet2009modeling} for a review on the modeling of nucleation processes. 

\par
\begin{figure}[h!]
  \begin{center}
    \includegraphics[width = .95\textwidth]{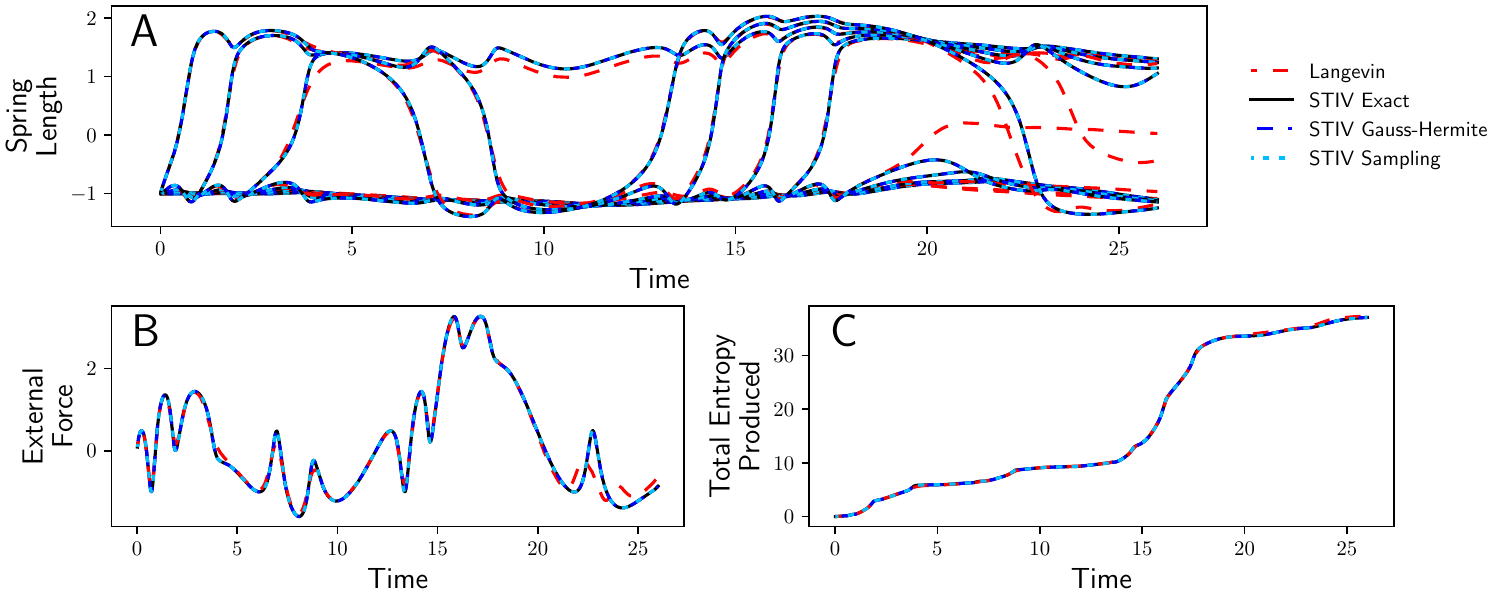}
     \caption{\label{fig:4new} A comparison of the average spring length (A), average external force (B), and total entropy produced (C) for the mass-spring-chain system with quartic interaction potential and non-linear external pulling protocol. The results are shown for Langevin simulations (red dashes), exact STIV (black solid), Gauss-Hermite STIV (blue dashed), and sample based STIV (light blue dotted). Throughout, the three \stiv methods collapse to a single line, indicating the accuracy of the two approximation techniques.}
  \end{center}
\end{figure}

In Fig.~\ref{fig:4new}, we show the same quantities as shown in Fig.~\ref{fig:4res} but for a new, cyclic, non-linear external pulling protocol, $\lambda(t) = l_0 + 5\sin(2\pi t/15) + t/2$ where $l_0$ is the unstretched length with all springs in the left well. This reveals another advantage of the \stiv method; the method does not need to be tuned to any specific pulling protocol to provide a good approximation. Although the approximation begins to break down at the end of the simulation, it is remarkable that the choice of a diagonal Gaussian approximation to the density of states produces predictions of this accuracy considering the multimodal nature of the true density of states induced by cyclical driving and repeated transitions.

\subsection{Coiled-coil protein phase transition}
\begin{figure}[h!]
  \begin{center}
    \includegraphics[width=.33\textwidth]{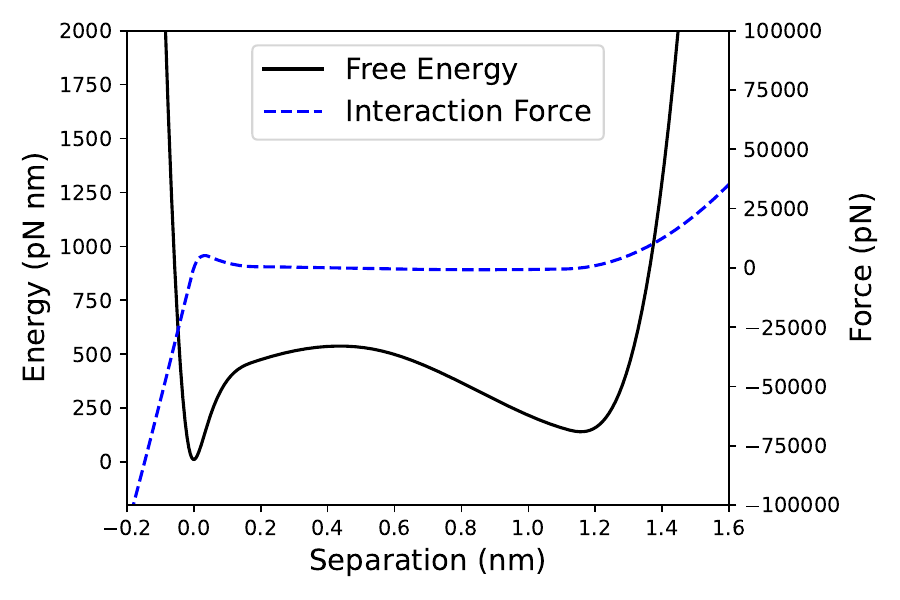}
    \caption{\label{fig:interaction}Coiled-coil interaction free energy (black solid line) and interaction force (blue dashed line). The original scatter point data for the free energy and interaction force comes from molecular dynamics simulations in \cite{torres2019}. These are linearly interpolated here.}
  \end{center}
\end{figure}

\begin{figure}[h!]
  \begin{center}
    \includegraphics[width=.95\linewidth]{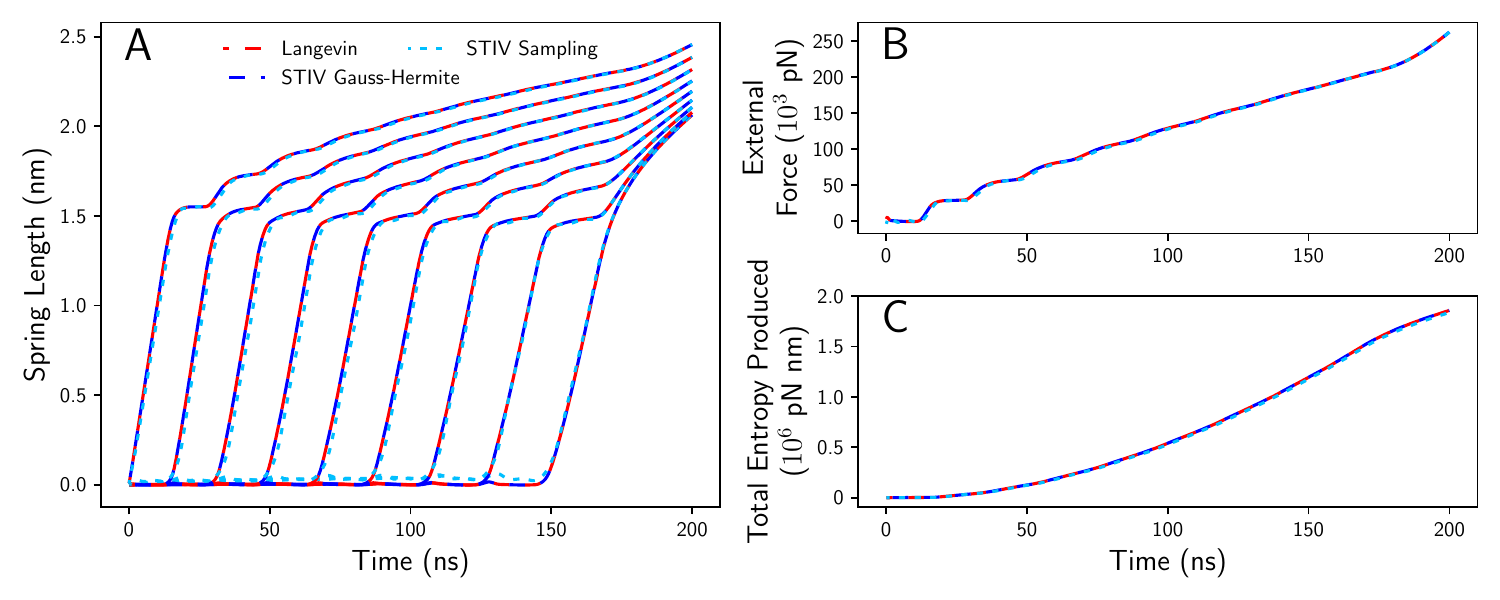}
    \caption{\label{fig:protein} A comparison of the average spring length (A), average external force (B), and total entropy produced (C) for the coiled-coil protein interaction potential and constant velocity external pulling. The results are shown for Langevin simulations (red dashes), Gauss-Hermite \stiv (blue dashed), and sample based \stiv (light blue dotted). }
  \end{center}
\end{figure}

As a final example, we apply the \stiv framework to study phase transformations in coiled-coil proteins. 
Under external loading, coiled-coils can transition form a tightly bound $\alpha$-helical state to an unfolded and elongated state \cite{torres2019}. 
This transition facilitates their ability to withstand large strains, is key to their biological function, and can be described using a double-well free energy landscape with mean length of the hydrogen bonds within the coiled-coil serving as the reaction coordinate. 
Here, we make use of the Gauss-Hermite and sampling based \stiv methods to study the behavior of a series of coiled-coils under external loading. 
In Fig.~\ref{fig:interaction}, we show the free energy landscape obtained from all atom molecular dynamics simulations \cite{torres2019}, which we take as the interaction energy in the mass-spring-chain configuration.
Since phase transitions are largely dominated by nucleation events at realistic viscosities ($\eta \approx 4$ cP), we use an artificially high viscosity to study the ability of \stiv to capture phase front propagation for this potential. 
The results for spring length, force and total entropy produced are shown in Fig.~\ref{fig:protein}, where we see good agreement between both numerical \stiv methods and Langevin simulations of the coiled-coil system. This example demonstrates that the STIV framework and suggested numerical implementations can be used with realistic interparticle potentials.

\section{Discussion} \label{sec:discussion}
We have shown that the \stiv framework  with a (strictly diagonal) Gaussian approximation of the density of states produces accurate models of phase transformation in one dimensional elastic systems driven by front propagation. The resulting models differ from traditional phase field models in two major ways: the phase field equations and equilibrium equations are coupled, and the non-equilibrium free energy driving the evolution is non-local but does not take the typical Ginzburg-Landau form. 
The \stiv framework has several advantages over traditional phase field: it requires no parameter tuning and hence does not need to be tuned to a specific external pulling protocol, it generates highly accurate predictions for thermodynamic quantities, and it can nucleate phase transformations due to external loading without additional physics for all examples tested. Moreover, it requires significantly less computational time to run and compute thermodynamic quantities than traditional Langevin simulations. 
However, since the \stiv model is designed to respect the physics at the microscopic scale, we remark that the phase field model induced by \stiv resolves the true interface width of the traveling phase front. For computational efficiency, most phase field models simulate a diffuse interface, and either show that model predictions are independent of the interface width or provide a way of correcting for the diffuse interface approximation. This being the case, the computational efficiency of the \stiv phase field model will likely scale more slowly than traditional phase field models for large systems and physically small interfaces. 

In addition, we have provided two methods for implementing the \stiv framework numerically, allowing one to approximate systems with arbitrary interaction energies. 
One method, based on Gauss-Hermite quadrature, is specific to the Gaussian approximation of the density of states. 
Although certainly restrictive, the Gaussian approximation to the density of states works remarkably well where external driving is the primary reason for phase transitions. This is in spite of the fact the true density of states is highly multimodal during transitions (when it is possible to find the system in one of two energy wells with non-zero probability). 
The second method, based on sampling, is generally less accurate for the same computational cost, but can be implemented for more general approximate density of states (any density which allows for easy sampling). The use of a more flexible approximate density of states, especially chosen for the system of study as a mechanism of coarse-graining, offers a potentially fruitful avenue of future study.

\section{Data Availability}
All data and source code used in generating this manuscript will be made available and linked here prior to final submission. 

\section*{Acknowledgments}
T.~L.~acknowledges that this project was supported in part by a fellowship award under contract FA9550-21-F-0003 through the National Defense Science and Engineering Graduate Fellowship Program, sponsored by the Army Research Office. P.~K.~P.~ acknowledges partial support through NIH grant R01 HL-148227. C.~R.~acknowledges support from NSF CMMI-2047506.

\section*{Disclosures}
The authors declare they have no conflicts of interest.

\section*{Author Contributions}
\textbf{Travis Leadbetter:} Conceptualization, Methodology, Software, Formal Analysis, Writing - Original Draft, Writing - Review \& Editing. 
\textbf{Prashant K. Purohit:} Writing - Review \& Editing, Supervision.
\textbf{Celia Reina:} Conceptualization, Methodology, Writing - Review \& Editing, Supervision.

\appendix

\section{Kinetic equations for a mass-spring-chain with a general interaction potential} \label{App:Derivation_KineticEquations}
We now provide the kinetic equations for the internal variables for a one dimensional mass-spring-chain with general interaction potential between particles and a Gaussian approximation of probability distribution with mean $\mu$ and diagonal covariance matrix with entries $\sigma^2$. Letting $u(z)$ be the arbitrary pairwise interaction potential as a function of the particle separation, so that the force on particle $i$ is $f_i(x) = f(x_i - x_{i-1}) - f(x_{i+1} - x_i)$ where $f(z) = -u'(z)$, and writing
\begin{equation*}
\hat{s} = -k_B\log(\hat{p}) = k_B\sum_{i=1}^N \left\{\frac{(x - \mu)_i^2}{2\sigma_i^2} + \frac1{2}\log(2\pi\sigma^2_i) \right\},
\end{equation*}
the dynamical equations for the internal variables can be found to be (SI Appendix section 3 of \cite{stiv})
\begin{equation}
 \avgph{\frac{\pd \hat{s}}{\pd \alpha_i}\frac{\pd \hat{s}}{\pd \alpha_j}} \dot{\alpha}_j= - k_B\avgph{ \fL^\dagger \frac{\pd \hat{s}}{\pd \alpha_i}},
\end{equation}
that is, for $\alpha=(\mu,\sigma)$,
\begin{align*}
\avgph{\frac{\pd \hat{s}}{\pd \mu_i}\frac{\pd \hat{s}}{\pd \mu_j}} \dot{\mu}_j + \avgph{\frac{\pd \hat{s}}{\pd \mu_i}\frac{\pd \hat{s}}{\pd \sigma_j}} \dot{\sigma}_j &= -k_B\avgph{ \left(\frac1{\eta}f_k \frac{\pd}{\pd x_k}  + d\frac{\pd}{\pd x_k}\frac{\pd}{\pd x_k} \right) \frac{\pd \hat{s}}{\pd \mu_{i}} } \\
\avgph{\frac{\pd \hat{s}}{\pd \sigma_i}\frac{\pd \hat{s}}{\pd \mu_j}} \dot{\mu}_j + \avgph{\frac{\pd \hat{s}}{\pd \sigma_i}\frac{\pd \hat{s}}{\pd \sigma_j}} \dot{\sigma}_j &= -k_B\avgph{ \left(\frac1{\eta}f_k \frac{\pd}{\pd x_k}  + d\frac{\pd}{\pd x_k}\frac{\pd}{\pd x_k} \right) \frac{\pd \hat{s}}{\pd \sigma_{i}} }.
\end{align*}
The left hand side matrices simplify to $\avgph{\frac{\pd \hat{s}}{\pd \mu_i}\frac{\pd \hat{s}}{\pd \mu_j}} = \delta_{ij}\frac{k_B^2}{\sigma_i^2}$, $\avgph{\frac{\pd \hat{s}}{\pd \sigma_i}\frac{\pd \hat{s}}{\pd \sigma_j}} = \delta_{ij}\frac{2k_B^2}{\sigma_i^2}$, and $\avgph{\frac{\pd \hat{s}}{\pd \sigma_i}\frac{\pd \hat{s}}{\pd \mu_j}}=0$, whereas for the right hand side we have 
\begin{align*}
f_k\frac{\pd}{\pd x_k}\frac{\pd \hat{s}}{\pd \mu_i} &= -\frac{k_B f_i}{\sigma_i^2} \\
\frac{\pd}{\pd x_k}\frac{\pd}{\pd x_k}\frac{\pd \hat{s}}{\pd \mu_i}  &= 0 \\
f_k\frac{\pd}{\pd x_k}\frac{\pd \hat{s}}{\pd \sigma_i} &= -\frac{2k_B f_i(x - \mu)_i}{\sigma_i^3} \\
\frac{\pd}{\pd x_k}\frac{\pd}{\pd x_k}\frac{\pd \hat{s}}{\pd \sigma_i}  &= -\frac{2k_B}{\sigma_i^3}.
\end{align*}
Hence,
\begin{align*}
\dot{\mu}_i &= \frac1{\eta}\avgph{f_i} \\
\dot{\sigma}_i &= \frac{1}{\eta}\avgph{\frac{f_i(x-\mu)_i}{\sigma_i}} + \frac{d}{\sigma_i}.
\end{align*}
If we define $\hat{f}(\epsilon,\tau)= \avg{-u'(z)}_{z\sim \N(\epsilon,\tau^2)}$ as in the main text, then by noting that $x_i - x_{i-1}$ is Gaussian with mean $\epsilon_i = \mu_i - \mu_{i-1}$ and variance $\tau_i^2 = \sigma_i^2 + \sigma_{i-1}^2$ under $\hat{p}$, we see that 
$\avgph{f_i} = \hat{f}(\epsilon_i,\tau_i) - \hat{f}(\epsilon_{i+1},\tau_{i+1})$ and we recover the equation for $\dot{\mu}$. As for the equation for $\dot{\sigma}_i$, we first note that 
\begin{align*}
\avgph{f(x_i - x_{i-1})(x - \mu)_i} &= \int f(x_i - x_{i-1})(x - \mu)_i \frac{\exp(-\frac{(x - \mu)_i^2}{2\sigma_i^2})}{\sqrt{2\pi\sigma_i^2}}\frac{\exp(-\frac{(x - \mu)_{i-1}^2}{2\sigma_{i-1}^2})}{\sqrt{2\pi\sigma_{i-1}^2}}\diff x_i\diff x_{i-1} \\
&=\sigma_i^2 \frac{\pd}{\pd\mu_i}\int f(x_i - x_{i-1})\frac{\exp(-\frac{(x - \mu)_i^2}{2\sigma_i^2})}{\sqrt{2\pi\sigma_i^2}}\frac{\exp(-\frac{(x - \mu)_{i-1}^2}{2\sigma_{i-1}^2})}{\sqrt{2\pi\sigma_{i-1}^2}}\diff x_i\diff x_{i-1} \\
&= \sigma_i^2 \frac{\pd}{\pd\mu_i} \hat{f}(\epsilon_i,\tau_i) \\
&= \sigma_i^2 \frac{\pd}{\pd \epsilon}\hat{f}(\epsilon_i,\tau_i).
\end{align*}
Likewise, $\avgph{f(x_{i+1} - x_{i})(x - \mu)_i} = -\sigma_i^2\frac{\pd\hat{f}}{\pd \epsilon} (\epsilon_{i+1},\tau_{i+1})$. 
Combining this with $d = 1/\eta\beta$ finally gives
\begin{equation*}
\dot{\sigma}_i = \frac{\sigma_i}{\eta}\left(\frac{\pd\hat{f}}{\pd \epsilon} (\epsilon_i,\tau_i) + \frac{\pd \hat{f}}{\pd \epsilon}(\epsilon_{i+1},\tau_{i+1}) + \frac1{\beta\sigma_i^2} \right).
\end{equation*}
Using the derivations from the next section (\ref{App:ThermodynamicsEquations}), we have $\frac{\pd\hat{E}}{\pd \sigma_i} = -\eta\dot{\sigma}_i + \frac1{\beta\sigma_i}$ and $T\frac{\pd\hat{S}^\tot}{\pd\sigma_i} = \frac1{\beta\sigma_i}$ so we can establish the gradient flow equation
\begin{equation*}
-\eta \dot{\sigma} = \frac{\pd\hat{A}^\noneq}{\pd\sigma}.
\end{equation*}
Similarly, as shown in \ref{App:ThermodynamicsEquations}, $\frac{\pd}{\pd \epsilon_i} \hat{A}^{\noneq}=-\hat{f}(\epsilon,\tau)$, and thus 
\begin{equation}
    \eta\dot{\mu} = - \frac{\pd \hat{A}^{\noneq}}{\pd \mu}.
\end{equation}

\section{Thermodynamics equations for a mass-spring-chain with a general potential} \label{App:ThermodynamicsEquations}
We start by computing the non-equilibrium free energy $\hat{A}^{\noneq}$ of a mass-spring-chain system with a general interaction potential when the probability distribution is approximated as Gaussian with mean $\mu$ and diagonal covariance matrix with entries $\sigma^2$. \stiv defines the non-equilibrium free energy as $\hat{A}^{\noneq} = \avgph{\hat{e} - T\hat{s}}$, and using $\hat{e} = e = \sum_{i=1}^{N+1} u(x_i - x_{i-1})$ ($x_0 \equiv 0$, $x_{N+1} \equiv \lambda$), $\hat{s} = -k_B\log(\hat{p})$, and $1/\beta = k_BT$, gives 
\begin{equation*}
\hat{A}^{\noneq} = \sum_{i=1}^{N+1}\hat{U}(\epsilon_i,\tau_i) - \frac1{2\beta}\sum_{i=1}^N \log(\sigma_i^2) - \frac{N}{2\beta}\left(1 + \log(2\pi)\right)
\end{equation*}
where $\hat{U}(\epsilon,\tau) = \avg{u(z)}_{z\sim \N(\epsilon,\tau^2)}$, $\epsilon_i = \mu_i - \mu_{i-1}$, and $\tau_i^2 = \sigma_i^2 + \sigma_{i-1}^2$ ($\mu_0 \equiv 0$, $\mu_{N+1}\equiv \lambda$, $\sigma_0 \equiv \sigma_{N+1} \equiv 0$). Next, we compute the work rate, $\frac{\diff}{\diff t} \hat{W} = \frac{\pd}{\pd\lambda} \hat{A}^{\noneq}\dot{\lambda}$.
We have
\begin{equation*}
\frac{\pd}{\pd \lambda} \hat{A}^{\noneq} = \frac{\pd}{\pd\epsilon} \hat{U}(\epsilon_{N+1},\tau_{N+1}) = -\hat{f}(\epsilon_{N+1},\tau_{N+1}),
\end{equation*}
where the second equality follows from
\begin{align*}
\frac{\pd}{\pd\epsilon} \hat{U}(\epsilon,\tau) &= \frac{\pd}{\pd\epsilon} \int u(z)\frac{\exp(-\frac{(z - \epsilon)^2}{2\tau^2})}{\sqrt{2\pi\tau^2}}\diff z \\
&= \int u(z)\frac{(z- \epsilon)}{\tau^2}\frac{\exp(-\frac{(z - \epsilon)^2}{2\tau^2})}{\sqrt{2\pi\tau^2}}\diff z\\
&= -\int u(z)\pd_z\frac{\exp(-\frac{(z - \epsilon)^2}{2\tau^2})}{\sqrt{2\pi\tau^2}}\diff z \\
&= \int u'(z)\frac{\exp(-\frac{(z - \epsilon)^2}{2\tau^2})}{\sqrt{2\pi\tau^2}}\diff z \\ 
&=-\hat{f}(\epsilon,\tau).
\end{align*}
Thus, $\frac{\diff}{\diff t}\hat{W} = -\hat{f}(\epsilon_{N+1},\tau_{N+1})\dot{\lambda}$. Moreover, this also shows that 
\begin{equation*}
\frac{\pd}{\pd\mu_i}\hat{E} = \frac{\pd}{\pd\epsilon} \hat{U}(\epsilon_i,\tau_i) - \frac{\pd}{\pd\epsilon} \hat{U}(\epsilon_{i+1},\tau_{i+1}) = -\hat{f}(\epsilon_i,\tau_i) + \hat{f}(\epsilon_{i+1},\tau_{i+1}) = -\eta \dot{\mu}_i.
\end{equation*}
On the other hand, 
\begin{equation*}
\frac{\pd}{\pd\sigma_i}\hat{E} = \frac{\sigma_i}{\tau_i}\frac{\pd}{\pd\tau}\hat{U}(\epsilon_i,\tau_i) + \frac{\sigma_i}{\tau_{i+1}}\frac{\pd}{\pd\tau}\hat{U}(\epsilon_{i+1},\tau_{i+1})
\end{equation*}
which we can simplify using 
\begin{align*}
\frac{\pd}{\pd\tau} \left(\frac1{\tau}\exp(-\frac{(z - \epsilon)^2}{2\tau^2}) \right)&= \left(\frac{(z-\epsilon)^2}{\tau^4} - \frac1{\tau^2}\right)\exp(-\frac{(z - \epsilon)^2}{2\tau^2}) \\
&= -\tau\frac{\pd}{\pd z}\frac{\pd}{\pd\epsilon} \frac1{\tau}\exp(-\frac{(z- \epsilon)^2}{2\tau^2}).
\end{align*}
So following the same steps as before 
\begin{align*}
\frac{\pd}{\pd \tau} \hat{U}(\epsilon,\tau) &= \frac{\pd}{\pd \tau} \int u(z) \frac{\exp(-\frac{(z - \epsilon)^2}{2\tau^2})}{\sqrt{2\pi\tau^2}} \diff z \\
&=\int \frac{u(z)}{\sqrt{2\pi}} \frac{\pd}{\pd \tau} \left(\frac1{\tau}\exp(-\frac{(z - \epsilon)^2}{2\tau^2})\right)\diff z\\
&= \int \frac{u(z)}{\sqrt{2\pi}} \left(-\tau \frac{\pd}{\pd z}\frac{\pd}{\pd \epsilon}\frac1{\tau}\exp(-\frac{(z - \epsilon)^2}{2\tau^2})\right) \diff z \\ 
&=\tau\frac{\pd}{\pd \epsilon}\int u'(z)\frac{\exp(-\frac{(z - \epsilon)^2}{2\tau^2})}{\sqrt{2\pi\tau^2}} \diff z \\
&=-\tau \frac{\pd}{\pd \epsilon}\hat{f}(\epsilon,\tau).
\end{align*}
Thus,
$\frac{\pd}{\pd\tau} \hat{U}(\epsilon_i,\tau_i) = -\tau_i \frac{\pd}{\pd\epsilon} \hat{f}(\epsilon_{i},\tau_{i})$ and $\frac{\pd}{\pd\tau} \hat{U}(\epsilon_{i+1},\tau_{i+1}) = -\tau_{i+1} \frac{\pd}{\pd\epsilon} \hat{f}(\epsilon_{i+1},\tau_{i+1})$, and we have 
\begin{equation*}
\frac{\pd}{\pd\sigma_i}\hat{E} = -\sigma_{i}\left(\frac{\pd}{\pd\epsilon} \hat{f}(\epsilon_i,\tau_i) + \frac{\pd}{\pd\epsilon} \hat{f}(\epsilon_{i+1},\tau_{i+1})\right) = -\eta\dot{\sigma}_i + \frac1{\beta\sigma_i}.
\end{equation*}
Thus, the heat rate is given by 
\begin{equation*}
\frac{\diff}{\diff t} \hat{Q} = -\eta \abs{\dot{\mu}}^2 - \eta \abs{\dot{\sigma}}^2 + \frac1{\beta}\sum_{i=1}^N \frac{\dot{\sigma}_i}{\sigma_i}.
\end{equation*}
Finally, the entropy production can be found via 
\begin{align*}
T\frac{\diff}{\diff t}\hat{S}^{\tot}  &= -\frac{\pd \hat{A}^{\noneq}}{\pd\mu} \cdot\dot{\mu} - \frac{\pd \hat{A}^{\noneq}}{\pd\sigma} \cdot\dot{\sigma} \\ 
&= -\frac{\pd\hat{E}}{\pd \mu}\cdot \dot{\mu} - \frac{\pd \hat{E}}{\pd \sigma}\cdot \dot{\sigma} + T \frac{\pd \hat{S}}{\pd \sigma} \cdot \dot{\sigma} \\
&= \eta\abs{\dot{\mu}}^2 + \eta\abs{\dot{\sigma}}^2
\end{align*}
as $T\hat{S} = \left(\sum_{i=1}^N \log(\sigma_i^2) + N(1 + \log(2\pi))\right)/2\beta$ and so 
\begin{equation*}
T\frac{\pd\hat{S}}{\pd \sigma_i} = \frac1{\beta\sigma_i}.
\end{equation*}

\section{Langevin simulations and numerical integration}
Throughout, we follow standard methods for solving stochastic and deterministic differential equations. A brief summary is provided here for clarity, and more details can be found in the SI Appendix of \cite{stiv}. 
\par
All stochastic Langevin equations of the form shown in Eq.~\eqref{eqn:Langevin} are solved using a first order Euler-Maruyama scheme with fixed time step. Likewise, ordinary differential equations of the form $\dot{\alpha}(t) = f(\alpha,t)$ for the dynamics of the internal variables and phase field variables are solved using a fourth order Runge-Kutta integration scheme with fixed time step. The time step varied between examples but was always chosen to be significantly smaller than the shortest effective relaxation time of the two potential minima, 
\begin{equation*}
\Delta t << \text{min}\left(\frac{\eta}{\abs{f'(\text{left minima})}},\frac{\eta}{\abs{f'(\text{right minima})}}\right).
\end{equation*}
Expectation values of thermodynamic quantities were determined from Langevin simulations using empirical averages from 10,000 sample trajectories, except in the case of the rate of total entropy production. In this case, the entropy of the true density of states was estimated using Gaussian kernel density estimation \cite{gkde}. 
\par 
For the quartic interaction potential, the thermodynamic and simulation parameters are number of masses $N = 8$, $\beta = 45$, $\eta = 0.4$, $l_1 = 1$, $l_2 = 1.5$, $a = 1$, and $\Delta t = 10^{-4}$. The Langevin simulation consisted of $10,000$ sample trajectories. Expectations for the Gauss-Hermite implementation of STIV were computed using the $16$ point Gauss-Hermite quadrature rule. The sampling based implementation of STIV used parameters of; Chebyshev points of order $N_\epsilon = N_\tau = 76$ on a grid of $\epsilon,\tau \in [-2,3]\times[0,0.35]$, the order of polynomial approximation $T_\epsilon = T_\tau = 6$, number of samples $N = 40,000$.  
For the coiled-coil protein interaction potential, the parameters are $N = 8$, $\beta = 0.2336$ pN nm ($1/k_BT$ for $T = 310$ K), $\eta = 2.92\times 10^5$ pN ns / nm$^2$, and $\Delta t = 10^{-4} ns$. The Langevin simulation consisted of $10,000$ sample trajectories. Expectations for the Gauss-Hermite implementation of STIV were computed using the $16$ point Gauss-Hermite quadrature rule. The sampling based implementation of STIV used parameters of Chebyshev points of order $N_\epsilon = N_\tau = 25$ on a grid of $\epsilon,\tau \in [-0.1,2.5]\times[0,0.02]$, the order of polynomial approximation $T_\epsilon = T_\tau = 9$, number of samples $N = 50,000$. In both examples, the sample based implementation of STIV was repeated with differing boundary box sizes to ensure that the trajectories remained with the interpolation range of the approximate polynomial.

\bibliographystyle{unsrt}
\bibliography{main}

\end{document}